\documentclass[prb,twocolumn,showpacs,amsmath,amssymb]{revtex4}
\usepackage{graphicx,color}
\usepackage{dcolumn}
\usepackage{bm}
\begin{document}
\title{Lattice effects on the spin dynamics in antiferromagnetic molecular rings}
\author{ Leonardo Spanu$^{1}$ and Alberto Parola$^{2}$}
\affiliation{
 $^1$ INFM and Dipartimento di
Fisica ``A.Volta", Universit\`a di Pavia, I-27100 Pavia, Italy \\
$^2$ Dipartimento di Fisica e Matematica, Universit\`a dell'Insubria, I-22100 Como, Italy }
\date{\today}
\begin{abstract}
We investigate spin dynamics in antiferromagnetic (AF) molecular rings at finite temperature
in the presence of  spin-phonon ({\it s-p}) interaction. We  derive a general expression for the spin 
susceptibility in the weak {\it s-p} coupling limit and then we focus on  the low-frequency behavior, 
in order to discuss a possible microscopic mechanism for nuclear relaxation in this class of magnetic 
materials. To lowest order in a perturbative expansion, we find that the susceptibility takes a Lorentzian 
profile and  all spin operators ($S^x$, $S^y, S^z$) contribute to spin dynamics at wave vectors $q\ne 0$. 
Spin anisotropies and local {\it s-p} coupling play a key role in the proposed mechanism. 
Our results prove that  small  changes in the spatial symmetry of the ring induce qualitative changes in the spin 
dynamics at the nuclear frequency, providing a novel mechanism for nuclear relaxation. 
Possible experiments are proposed.
\end{abstract}
\pacs{75.50.Xx, 76.60.-k }
\maketitle
In a crystal of molecular magnets not all physical properties and microscopic interactions can be exhaustively 
described by an effective Hamiltonian containing only magnetic terms. The works on magnetization tunneling 
in Mn$_{12}$ compounds ~\cite{mn12,mertes,park} clearly demonstrated that the localized spins in each molecule 
interact with lattice degrees of freedom, dislocations ~\cite{mertes,park,dislocation} and organic 
ligand  deformations ~\cite{villain}, changing local field at each magnetic site. Spin-lattice interactions revealed 
thus an essential ingredient for explaining the relaxation of the macroscopic magnetization in ferromagnetic clusters.
In the case of antiferromagnetic (AF) rings, thermodynamic properties  such as specific heat and magnetization curves 
can be rather satisfactorily described in terms of a minimal $N-$spins Hamiltonian containing only magnetic interactions  
\begin{equation}
\label{hamispin}
H_{\sigma}=J \sum_{j=1}^{N} \vec{S}_{j}\cdot\vec{S}_{j+1} + D\,\sum_{j=1}^{N} 
(S_{j}^{z})^2 + g\mu_B \sum_{j=1}^{N}  \vec{H}\cdot\vec{S}_{j}
\end{equation}
where periodic boundary conditions are understood. 
In addition to the dominant Heisenberg interaction and to the Zeeman coupling, the anisotropy term in Eq. (\ref{hamispin})
($D\sim -\,10^{-2}\, J$) is an effective description of the local field at each magnetic site.  
This model has been also successfully adopted for interpreting the behavior of dynamical observables at the electronic 
frequency - electron paramagnetic resonance (EPR) \cite{cornia} or inelastic neutron scattering (INS) experiments \cite{cr8_neut}. 
However Hamiltonian (\ref{hamispin}) proves not adequate for explaining nuclear relaxation phenomena in AF rings 
and a coupling with external degrees of freedom is then required: The discrete spectrum of the molecular magnet does not allow 
the relaxation between Zeeman-split nuclear levels, due to the mismatch between the typical nuclear energy scale 
($10^{-5} J $) and the  electronic one (ranging from $J$ to $g \mu_b H\sim 10^{-2} \,J$). NMR experiments performed 
on AF rings ~\cite{borsarev} suggest the relevance of the spin-phonon ($s-p$) coupling for understanding and explaining 
the relaxation rate $1/T_1$  ~\cite{borsarev,ferro6,baek} and, more generally, the spin dynamics at very low frequency in 
these systems. 
In spite of the large amount of experimental NMR data available in Fe$_6$ ~\cite{ferro6} and Cr$_{8}$ ~\cite{baek}, 
a comprehensive microscopic description of the relaxation mechanism in AF rings at finite temperature and weak magnetic 
field is still missing. Past investigations were mainly focused on the dynamics at high magnetic field and 
zero temperature in connection to the problem of tunneling of the N\'eel vector ~\cite{loss_tunnel}.
Assuming that the Hamiltonian (\ref{hamispin}) is a good starting effective model for the AF rings, a central 
problem is to clarify which spin mechanisms control the low-frequency dynamics, allowing to observe the Lorentzian 
profile of the relaxation rate $1/T_1$  and its  strong frequency dependence ~\cite{borsarev,baek}. Moreover, 
it is not fully understood which magnetic property may be investigated by the nuclear relaxation measurements, 
due to the mismatch between the nuclear and the electronic energy scale. 
A microscopic description of the  {\it s-p}  effects on the spin dynamics should also help the 
identification of the decoherence mechanism in this class of materials, the interaction of the spins with external 
degrees of freedom strongly limiting possible future technological applications ~\cite{loss_computer}.

In this Letter we investigate the dynamical spin susceptibility in the presence of a spin-phonon coupling in order to 
identify the possible microscopic spin mechanisms at the basis of the observed dynamics. We derive a 
general expression for the dynamical spin correlation functions and then  we focus on  the very low frequency limit, 
connecting our results to the problem of the nuclear relaxation in AF rings. 
We will show that, to lowest order in  the {\it s-p} coupling, even small perturbations breaking the 
reflection symmetry of the molecular ring have a profound effect on the relaxation rate, due to 
contributions coming from the dynamics of all spin operators ($S^z$ and $S^x$,$S^y$) at wave vectors $k\ne 0$. 
In a recent work ~\cite{santini} based on a Master Equation formalism, {\it Santini et al.} found that in the 
$\omega\to 0$ limit, the spin dynamics at low temperature, in the presence of a coupling with a thermal bath, is dominated by 
the $S^z$ operator at $q=0$ wave vector. As clearly shown in a related context, however, 
a fully microscopic approach is often essential for clarifying quantum effects in mesoscopic spin-phonon systems ~\cite{nostrum}. 

Due to  {\it s-p} coupling, although the nucleus does not directly interact with phonons ~\cite{quadrupolare}, 
the nuclear relaxation rate $1/T_1$  probes  the spin dynamics of the many-body system, 
the relaxation rate being expressed as ~\cite{rigamonti}
\begin{eqnarray}
\label{tuno}
\frac{1}{T_1}\propto\sum_{\sigma \tau} \sum_q ( A^{\sigma \tau}_{q} S^{\sigma \tau}(q,\omega_L) + 
(A^{\sigma \tau}_{q})^{*}S^{\sigma \tau}(q,-\omega_L))
\end{eqnarray}
In Eq. (\ref{tuno}) $A^{\sigma \tau}_{q}$ are geometrical coefficients and $S^{\sigma \tau}(k,\omega)$ 
represents the dynamical spin correlation function at the nuclear Larmor frequency $\omega_L \sim 10^{-5} J$. 
The dynamical correlations may be conveniently expressed in Lehmann representation as 
\begin{eqnarray}
\label{leman}
S^{\sigma\tau}(k,\omega)&=&\frac{1}{N Z_T} \sum_{i,f} \, e^{-\beta E_i}\,
<i|S_{k}^{\sigma}|f><f|S_{-k}^{\tau}|i> \nonumber\\
&\times&\delta (E_i-E_f-\omega)
\end{eqnarray}
where $S^{\sigma}_q$ is a generic single spin operator ($k=2 \pi n/N$ $n=0 \dots N-1$), and
$|i>$ and $|f>$ are the many-body exact eigenstates of the full Hamiltonian, 
including spin and phonon degrees of freedom, with energy $E_i$ and $E_f$ respectively.
The system of interest is described by the minimal model Hamiltonian
\begin{eqnarray}
\label{hami}
H&=&H_{\sigma}+H_P + H_{sp} \\
H_P&=&\sum_q \omega_q a^\dagger_q a_q  \\
H_{sp}&=& V(\vec{S}) {1\over\sqrt{N}} \sum_q {1\over \sqrt{\omega_q}} (a_q+a^\dagger_q) 
\end{eqnarray}
where $a_q$ are the Bose operators for the (three dimensional) phonons, $\omega_q$ the phonon frequency 
and $V(\vec{S})$ is a generic 
spin operator describing the coupling with the phonons ({\it s-p} coupling constant is implicitly included in the 
definition of the potential $V(\vec{S})$). Assuming that the {\it s-p} term weakly perturbs the spin system, 
we express the generic state $|i>$ (and $|f>$) in Eq. (\ref{leman}) as a product of eigenstates $|\sigma>|m>$,  $|\sigma>$ 
describing the spins and $|m>$ the phonon degrees of freedom. To lowest order in a perturbative expansion in the 
{\it s-p} coupling, we take $|\sigma>$ as an exact eigenstate of the Hamiltonian (\ref{hamispin}). 
Then we write the eigenvalue equation for the state $|i>=|\sigma>|m>$ and, after projecting onto the spin state $|\sigma>$, 
we finally obtain the equation defining the phonon state $|m>$: 
\begin{eqnarray}
\label{perturba}
&&E_\sigma |m> + \sum_q \omega_q a^\dagger_q a_q |m> + \\
&&{<\sigma | V | \sigma> \over \sqrt{N}} 
\sum_q {1\over \sqrt{\omega_q}} (a_q+a^\dagger_q) |m> = E|m> \nonumber
\end{eqnarray}  
The exact solutions of Eq. (\ref{perturba}) are products over $q$ of shifted harmonic oscillator states, 
the shift constant $\alpha_q^\sigma$ depending on the form of the {\it s-p} coupling $V(\vec S)$ and on the spin state: 
$\alpha^\sigma_q = {<\sigma | V | \sigma> \over \sqrt{N\omega_q^3}}$. Substituting the product eigenstates into 
the formal expression (\ref{leman}) we obtain
\begin{eqnarray}
\label{general1}
S^{\sigma \tau}(k,\omega)&=&\frac{1}{Z} \sum_{\sigma_i,\sigma_f}\sum_{n,m}  
e^{-\beta (E_i + \sum_q \omega_q n_q)} |<n|m>|^2 \nonumber\\
&\times&<\sigma_i|S_{k}^{\sigma}|\sigma_f><\sigma_f|S_{-k}^\tau|\sigma_i> \nonumber\\
&\times&\delta (\Delta E_{if} +\sum_q \omega_q (n_q-m_q)-\omega)
\end{eqnarray}  
where $|\sigma_i>$ and $|\sigma_f>$ identify pure spin states with  energy difference $\Delta E_{if}$ and  
$|n>$ and $|m>$ are the phonon states with phonon occupation number $\{n_q\}$ and $\{m_q\}$ respectively. 
The key element in Eq. (\ref{general1}) is the overlap between the two shifted harmonic oscillator 
eigenstates $|<n|m>|$, the two shifts depending on the spin states $|\sigma_i>$ and $|\sigma_f>$. 
Using an integral representation of the delta function in Eq. (\ref{general1}) and the explicit expression for 
$|<n|m>|$ \cite{landau} we obtain
\begin{eqnarray}
S^{\sigma \tau}(k,\omega) &=& \frac{Z_p}{Z} \sum_{\sigma_i,\sigma_f}e^{-\beta E_i}
<\sigma_i|S_{k}^{\sigma}|\sigma_f> \nonumber \\
&& \times <\sigma_f|S_{-k}^\tau|\sigma_i>\,I_{if}(\omega) 
\end{eqnarray}
where $Z_p$ is the phonon partition function and the function $I_{if}(\omega)$ is defined (see also ~\cite{legget})
\begin{eqnarray}
I_{if}(\omega)&=&
\int {dt\over 2\pi}
e^{i(\Delta E_{if} -\omega)t} \times \\
&& e^{-\sum_q (\Delta \alpha_q)^2
[i\sin\omega_q t +\coth(\beta\omega_q/2)(1-\cos\omega_q t)]} \nonumber 
\label{integrale}
\end{eqnarray}   
In the simple situation of a phonon spectrum of the Debye type: $\omega_q = c q$ for $\omega< \omega_D$, 
taking the thermodynamic limit and considering the low temperature behavior $\beta \omega_D \to \infty$, 
Eq. (\ref{integrale}) becomes
\begin{eqnarray}
\label{integrale2}
I_{if}=\frac{\beta}{2} 
\int_{-\infty}^{\infty} {dy\over 2\pi} \,e^{-i\Omega_{if} y}\, 
e^{-\gamma \Lambda(\frac{1}{2}\beta\omega_D y)}\,
\left [\frac{\sinh \frac{\pi}{2} y} {\frac{\pi}{2} y}\right ]^{-\gamma_{if}}
\end{eqnarray}
where $\Lambda(u)=\int_{0}^{u}dz\frac{1-e^{iz}}{z}$ and we have introduced 
two important quantities, the dimensionless frequency $\Omega_{if}=\frac{1}{2}\beta(\omega - \Delta E_{if})$
and the effective coupling 
\begin{equation}
\label{gamma}
\gamma_{if}=3\,\omega_D^{-3}\,[<i|V|i>-<f|V|f>]^2
\end{equation}
In the  weak {\it s-p}  coupling limit, $\gamma_{if} \to 0$ and Eq. (\ref{integrale2}) reduces to 
\begin{eqnarray}
I_{if}(\omega)\sim \frac{\beta\gamma_{if}} {\pi^2\gamma_{if}^2 +4\Omega^2} 
\label{ultima}
\end{eqnarray}
Finally, factorizing the partition function in terms of phonon and spin terms $Z=Z_pZ_\sigma$ 
and collecting together the previous expressions, we obtain  
\begin{eqnarray}
\label{ultima2}
S^{\sigma \tau}(k,\omega)&=&\frac{1}{Z_{\sigma}} \sum_{\sigma_i,\sigma_f}<\sigma_i|S_{k}^{\sigma}|\sigma_f>
<\sigma_f|S_{-k}^\tau|\sigma_i> \nonumber \\
&\times & \frac{ \gamma_{if}}{\beta(\pi^2 \gamma_{if}^2/\beta^2 + (\omega-\Delta E_{if})^2)}\, e^{-\beta E_i}
\end{eqnarray}
Eq. (\ref{ultima2}), valid in the weak coupling regime, reproduces a Lorentzian profile and allows to connect in 
a non-obvious way the three different energy scales present in our problem: the nuclear, the thermal and the magnetic one. 
Regardless the functional form of the spin-phonon interaction  $V(\vec{S})$, we first observe that 
when the initial spin state $|\sigma_i>$ is equal to the final one $|\sigma_f>$, the effective coupling $\gamma_{if}$ vanishes and 
$I_{if}$ reduces to a delta function at $\omega=0$. Within our formalism, the first non-zero contribution 
at low frequency in Eq. (\ref{ultima2}) is obtained by summing over pairs of spin states for 
which $\gamma_{if} \ne 0$. If the energy gap between the initial and final states is larger than
the frequency scale (i.e. $\Delta E_{if}>> \omega$), frequency dependence in Eq. (\ref{ultima2}) can be neglected, 
giving thus an $\omega-$independent relaxation rate $1/T_1$. The experimental observation of a 
strong frequency dependence in the $1/T_1$ suggests that other channels contribute to the relaxation  mechanism of the 
nuclear spin. Processes involving states originally belonging to the same spin multiplet are not good candidates, 
because  the energy splitting created by the presence of both anisotropy and field in a generic orientation is two or 
three orders of magnitude larger than the nuclear energy scale. In order to preserve a frequency dependence in 
Eq. (\ref{ultima2}) we are instead forced to consider transitions between pairs of quasi-degenerate spin states 
$E_{if}\sim \omega_L$. By examining the spectrum of Hamiltonian (\ref{hamispin}), we found that degenerate states connected by 
reflection symmetry - with momenta $q$ and $-q$ - are the only possible candidates for describing 
the mechanism governing the very low frequency dynamics at temperature $T\sim J$, within our approach.
\begin{figure}
\vspace{1mm}
\includegraphics[width=0.40\textwidth]{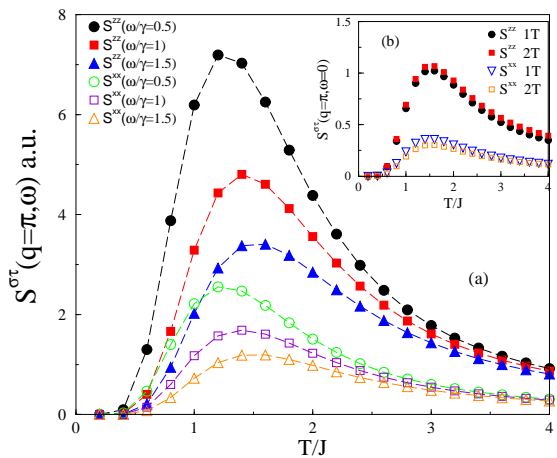}
\caption{\label{fig1}
Dynamical susceptibility  $S^{\alpha \beta}(q=\pi,\omega)$ as a function of the temperature $T/J$ for 
Cr$_8$ ($N=8$; $S=3/2$; $J\sim11.7 K$) calculated according to Eq. (\ref{ultima2}) at fixed magnetic field 
($h=1$ T,$\theta=0.5$ rad) and at different  frequencies (in units of $J$). Filled symbols refer to the $S^z$ operator, 
empty to $S^x$. Inset: Exact Diagonalization Results for the spin susceptibility of the pure spin system 
(i.e. no spin-phonon coupling) at $\omega=0$, calculated at two different magnetic fields. 
}      
\end{figure}
However, in order to get a non vanishing $\gamma_{if}$ via Eq. (\ref{gamma}), the {\it s-p} coupling operator
$V(\vec S)$ must have a non vanishing matrix element between the two states $|q>$ and $|-q>$, i.e. the 
reflection symmetry of the ring must be explicitly broken by the spin-phonon term. Starting from the Hamiltonian (\ref{hamispin}),
the interaction of each single ion with the neighboring organic atoms  provides the required perturbation terms. 
Actually it is well established that ligand groups 
surrounding each magnetic site can assume different, energetically equivalent, spatial orientations (quenched disorder), 
lowering the symmetry  of the molecule with respect to that of a perfect coplanar ring ~\cite{cr8_neut}. 
The positions of the atoms determine the strength and the type of interaction of each spin with the phonons, 
the spin-phonon potential being described in a quite general way by the expression\begin{equation}
\label{potenziale}
V=\sum_{i=1,N}C_i O(\vec{S}_i). 
\end{equation}
The operators $O(\vec{S}_i)$ describe electrical quadrupole interaction \cite{abragam,santini}, while $C_i$ is the 
site dependent {\it s-p} coupling which breaks the reflection symmetry of the Hamiltonian (\ref{hamispin}). 
A detailed description of the functional form of the quadrupolar operators $O(\vec{S}_i)$ and a quantitative estimate 
of each local spin-phonon coupling is beyond the aims of our work and it is not essential for a qualitative 
understanding of the proposed mechanism for the spin dynamics. The main point we want to stress is the existence 
of a site dependent interaction, breaking the translation and reflection symmetry of the ring: the degeneracy 
between the two states $|q>$ and $|-q>$ is lifted and a non vanishing coupling constant $\gamma_{if}$ (\ref{gamma})
is generated. The nuclear relaxation rate thus acquires a contribution from the spin dynamics at wave vector $k=2q$ showing 
a strong frequency dependence when the splitting created by the perturbation is comparable with the nuclear energy scale. 
The  differences in the interactions at each magnetic site can be usually neglected at energies higher than the 
nuclear one and site dependent parameters do not improve the quality of the experimental fit~\cite{cr8_neut}. 
On the contrary, we argue that, at the nuclear frequency,  very small differences in the {\it s-p} coupling are relevant 
for the relaxation rate $1/T_1$.  We find here a typical aspect of the disorder in correlated systems: 
macroscopic observables, as the $1/T_1$, are strongly affected by very small  perturbations, when the perturbation 
itself becomes comparable with  the main energy scale of the problem. 

\begin{figure}
\vspace{1mm}
\includegraphics[width=0.40\textwidth]{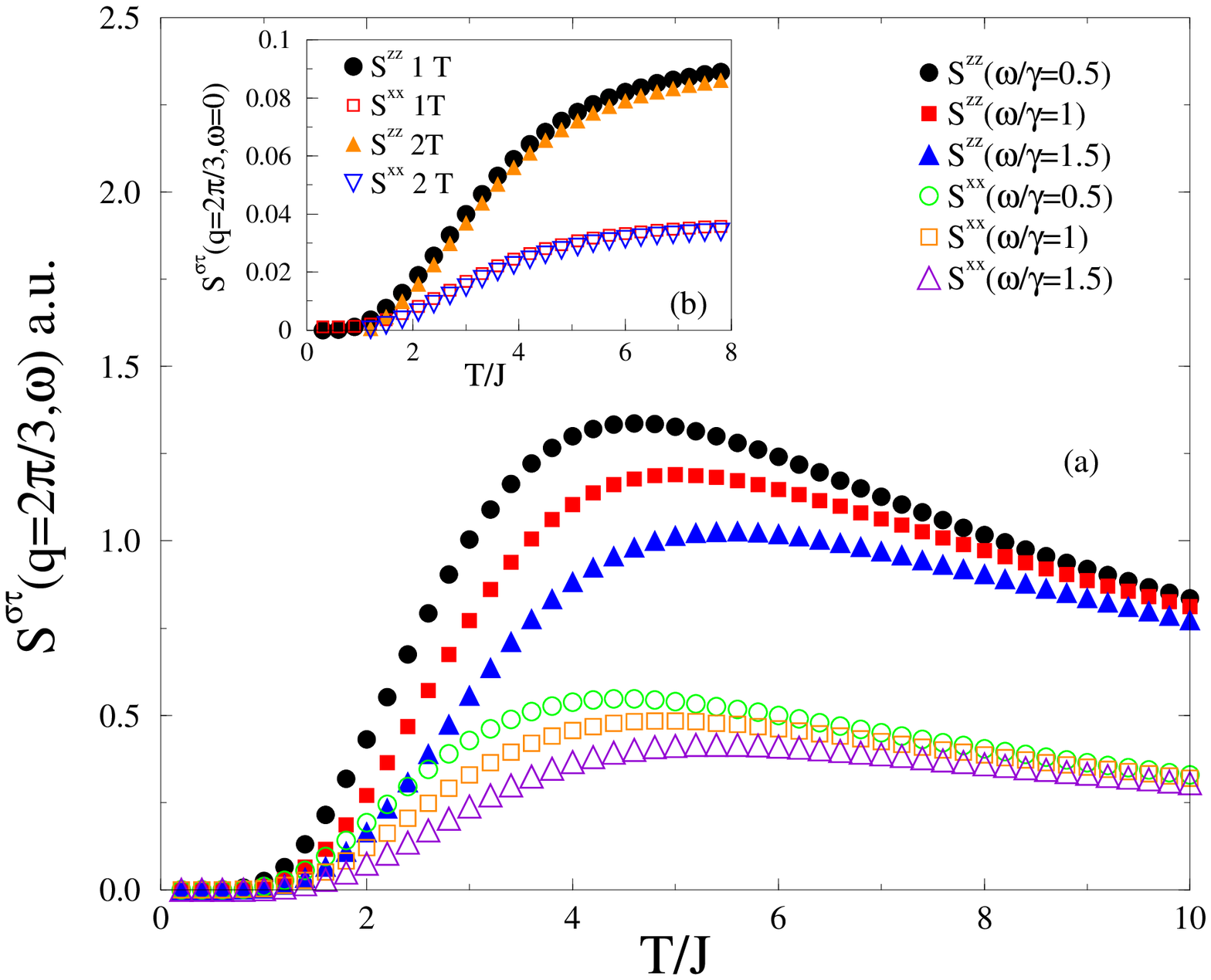}
\caption{\label{fig2}
Same as Fig. \ref{fig1} for Fe$_6$ ($N=6$; $S=5/2$; $J \sim28 K$). 
}
\end{figure}

Exact diagonalizations have been performed in order to evaluate the low frequency behavior of the dynamic correlations 
in Cr$_8$ and Fe$_6$ molecular rings. In Fig \ref{fig1} and in Fig. \ref{fig2} we report the results for 
the dynamical susceptibility $S^{\sigma,\tau}(k,\omega)$ as a function of temperature $T/J$ from Eq. (\ref{ultima2}), 
for the Cr$_8$ ($k=\pi$) and Fe$_6$ ($k=2\pi/3$) compounds. Calculations are performed at fixed 
magnetic field ($h=1$ T, in the direction $\theta=0.5$ rad) and for different frequencies, including in Eq.(\ref{ultima2}) 
pairs of quasi-degenerate states. The dynamical susceptibility of  both $S^z$ and  $S^x$ operators shows a 
maximum as a function of the temperature, the value of the maximum decreasing and its position moving to higher 
temperatures when the frequency is increased. This behavior qualitatively  reproduces the experimental observations in 
both materials \cite{borsarev}. The matrix elements in  Eq. (\ref{ultima2}) are only weakly dependent on the magnetic field, 
as can be inferred from the results for the spin susceptibility $S(k,\omega=0)$ of the pure spin Hamiltonian 
(insets of Fig \ref{fig1} and Fig. \ref{fig2}). 
Our results support the conclusion that the external magnetic field affects the relaxation rate 
mainly through the Larmor frequency $\omega_L=\Gamma_L \,H$ \cite{borsarev}.
In conclusion we investigated low-frequency spin dynamics at finite temperature in AF rings in the presence of 
{\it s-p} coupling, deriving a general expression for the dynamical susceptibility. We emphasized the central role of 
the lattice degrees of freedom, showing that the interaction of the spins with the surrounding atoms can generate 
a contribution to the dynamics at the nuclear frequency from non-diagonal operators $S^{\pm}_{2q}$, $q\ne0$. 
Small changes in the spatial symmetry of the ring induce thus novel effects in the spin dynamics at low frequency. 
A characteristic feature of our expressions is that the low frequency dynamics acquires a non-zero contribution 
from all spin operators  $S^z$ and $S^{\pm}$.
For isotropic hyperfine coupling $A^{\sigma\tau} \propto \delta^{\sigma\tau}$ (uniform magnetic field along 
$z-$direction, radio-frequency field in $x-$direction) $1/T_1$ involves only matrix elements of $S^{\pm}$.  
NMR experiments performed on Cr nucleus  in Cr$_8$ or on $^{57}$Fe in Fe$_6$, should provide an interesting 
test for detecting the contribution coming from transverse spin fluctuations at $q\ne0$. Conversely, the proposed 
mechanism is excluded or strongly suppressed when the reflection symmetry is broken at an energy scale larger
than the nuclear one. By applying hydrostatic pressure it is in principle possible to modify the global 
symmetry of  each molecule, verifying the role of the reflection symmetry breaking on the nuclear relaxation.

We gratefully acknowledge  F.Borsa and A. Lascialfari for valuable comments and important suggestions. 
We thank A. Rigamonti, P. Caretta, P. Santini and G. Amoretti for useful discussions.


\begin{thebibliography}{99}
\bibitem{quadrupolare} The absence of a quadrupolar interaction does not allow a direct coupling between the nuclear spin and the phonon; see for instance ref. ~\cite{abragam}.
\end{thebibliography}
\end{document}